\documentclass[nologo,11pt,a4paper]{ETHpaper}

\usepackage[pdftex]{graphicx}
\usepackage{subfigure}
\usepackage{amsmath, amssymb}
\usepackage{enumitem}
\usepackage{color, colortbl}
\usepackage{afterpage}
\usepackage{url}
\usepackage{multirow}
\usepackage{rotating}
\usepackage{scalefnt}
\usepackage{hyperref}

\begin{document}

\title{Causality-Driven Slow-Down and Speed-Up of Diffusion in Non-Markovian Temporal Networks}

\author{Ingo Scholtes$^*$, Nicolas Wider, Ren\'{e} Pfitzner, Antonios Garas, Claudio J. Tessone, Frank Schweitzer}
\authoralternative{Ingo Scholtes et al.}
\address{Chair of Systems Design, ETH Zurich \\
CH-8092 Zurich, Switzerland \\
$^*$\texttt{ischoltes@ethz.ch}}
\reference{preprint of article published in Nature Communications, Vol. 5, Article 5024, September 24, 2014}
\www{\url{http://www.sg.ethz.ch}}

\maketitle

\begin{abstract}
Recent research has highlighted limitations of studying complex systems with time-varying topologies from the perspective of static, time-aggregated networks.
Non-Markovian characteristics resulting from the ordering of interactions in temporal networks were identified as one important mechanism that alters causality, and affects dynamical processes.
So far, an analytical explanation for this phenomenon and for the significant variations observed across different systems is missing.
Here we introduce a methodology that allows to analytically predict causality-driven changes of diffusion speed in non-Markovian temporal networks.
Validating our predictions in six data sets, we show that - compared to the time-aggregated network - non-Markovian characteristics can lead to both a slow-down, or speed-up of diffusion which can even outweigh the decelerating effect of community structures in the static topology.
Thus, non-Markovian properties of temporal networks constitute an important additional dimension of complexity in time-varying complex systems.
\end{abstract}

\section{Introduction}

Complex systems in nature, society and technology are rarely static but typically have time-varying network topologies.
The increasing availability of high-resolution data on time-stamped or time-ordered interactions from a variety of complex systems has fostered research on how different aspects of the temporal dynamics of networks influence their properties.
Focusing on one particular aspect, a first line of research has studied the concurrency and duration of interactions~\cite{Morris1995,Aurell2009,Tessone2012,Masuda2013,Ribeiro2013}.
Some of these works show that compared to systems where, similar to static networks, most or all links are available concurrently, dynamical processes like epidemic spreading or diffusion are slowed down by the continuously switching topologies of temporal networks~\cite{Morris1995,Masuda2013,Ribeiro2013}.
Other works show that the dynamics of network topologies can introduce noise which fosters certain types of consensus processes~\cite{Aurell2009,Tessone2012}.
Assuming that network topologies change in response to the dynamical process running on top of it, another line of research has studied adaptive networks, again highlighting that network dynamics have important consequences for dynamical processes~\cite{Gross2006,Gross2009}.
Considering interactions in dynamic networks as a time series of events, a number of recent works focused on the question of whether observed inter-event times are consistent with the Poissonian distribution expected from a memoryless stochastic process.
For a number of dynamic social systems, it has been shown that inter-event times follow non-Poissonian, heavy-tail distributions, and that the resulting bursty interaction patterns influence the speed of dynamical processes like spreading and diffusion~\cite{Iribarren2009,Karsai2011,Rocha2011,Starnini2012,Perra2012,Perra2012a,Hoffmann2013,Takaguchi2013,Rocha2013b,Karsai2014,Jo2014}.
While all of these works highlight the importance of temporal information in the study of networks, there are a number of questions that have not been answered satisfactorily.
Most empirical studies of dynamical processes in temporal networks focus on the influence of heavy-tail inter-event time distributions in dynamic social networks, which likely result from human task-execution mechanisms~\cite{Grinstein2008,Garas2012,Jo2012}.
However, inter-event time distributions cannot explain temporality effects in other types of dynamic complex systems in which interactions are distributed homogeneously in time.
Furthermore, this approach requires that sufficiently precise time stamps can be assigned to interactions, thus excluding path-based data where merely the ordering of interactions can be inferred.

While inter-event time statistics have been studied in much detail, an important additional characteristic of temporal networks is that the \emph{ordering} of interactions influences \emph{causality}.
Different from static networks, the presence of two time-stamped edges $(a,b)$ and $(b,c)$ in a temporal network does not necessarily imply the existence of a path $a \rightarrow b \rightarrow c$ connecting node $a$ to $c$ via $b$.
Instead, so-called \emph{time-respecting paths} must additionally respect causality, i.e. a time-respecting path only exists if edge $(a,b)$ occurs \emph{before} edge $(b,c)$~\cite{Kempe2002,Holme2012}.
In order to additionally consider the \emph{timing} of interactions, it is common practice to impose the additional constraint that edges $(a,b)$ and $(b,c)$ must occur within a certain time window, thus imposing a limit on the time a particular process can wait in node $b$.
As such, both the \emph{order and timing} of interactions affect time-respecting paths - and thus causality - in temporal networks.
Compared to the rich literature on node activities, a relatively smaller number of studies empirically investigated effects of causality in temporal networks~\cite{Kempe2002,Kostakos2009,Kovanen2011,Rocha2013,Lentz2013,Pfitzner2013,Rosvall2013,Lambiotte2014}.
Recent works have shown that order correlations in temporal networks lead to \emph{causality structures} which significantly deviate from what is expected based on paths in the corresponding time-aggregated networks~\cite{Pfitzner2013,Lentz2013,Rosvall2013}.
Studying time-respecting paths $a \rightarrow b \rightarrow c$ from the perspective of a contact sequence $a,b,c$ passing through node $b$, it was shown that the next contact $c$ not only depends on the current contact $b$, but also on the previous one~\cite{Pfitzner2013,Rosvall2013,Lambiotte2014,Sun2014}.
As a consequence, contact sequences in real-world temporal networks exhibit \emph{non-Markovian characteristics} that are in conflict with the \emph{Markovian} assumption implicitly made when studying temporal networks from the perspective of time-aggregated networks, and which can neither be attributed to inter-event time distributions, nor to the concurrency or duration of interactions~\cite{Lentz2013,Pfitzner2013,Rosvall2013,Lambiotte2014}.
Furthermore, it was shown that \emph{causality structures} resulting from non-Markovian contact sequences influence both the speed of and the paths taken by dynamical processes~\cite{Pfitzner2013,Rosvall2013}.
These works not only question the applicability of the static network paradigm when modeling dynamic complex systems, they also highlight a \emph{temporal-topological} dimension of temporal networks which is ignored when exclusively focusing on time distributions of events and associated changes in the \emph{duration} of dynamical processes.
In line with the general lack of analytical approaches to understand and predict the effects of network dynamics on dynamical processes~\cite{Masuda2013,Porter2014},
an analytical explanation for the influence of causality structures in real-world complex systems, as well as for the significant variations observed across different systems, is currently missing.

To fill these gaps, in this article we introduce an analytical approach that allows to study dynamical processes in non-Markovian temporal networks.
In particular, we introduce higher-order time-aggregated representations of temporal networks that preserve causality, and use them to define Markov models for non-Markovian interaction sequences.
We show that the eigenvalue spectrum of the associated transition matrices explains the slow-down and speed-up of diffusion processes in temporal networks compared to time-aggregated networks.
We derive an analytical prediction for direction and magnitude of the change in a temporal network, validate it against six empirical data sets, and show that order correlations can both slow-down or speed-up diffusion even in systems with the same static topology.
Our results highlight that non-Markovian characteristics of temporal networks can either enforce or mitigate the influence of topological properties on dynamical processes.
As such, they constitute an important additional dimension of complexity that needs to be taken into account when studying time-varying network topologies.

\section{Results}

\subsection{Causality-driven changes of diffusive behaviour}
We define a \emph{temporal network} to be a set of directed, time-stamped edges $(v,w;t)$ connecting node $v$ to $w$ at a discrete time step $t$.
In this framework we assume time-stamped interactions $(v,w;t)$ to be instantaneous, occurring at time $t$ for exactly one discrete time step.
However interactions lasting longer than one time step can still be represented by multiple interactions occurring at consecutive time steps.
We further define a \emph{time-aggregated}, or \emph{aggregate}, network to be a projection along the time axis, i.e. a directed edge $(v,w)$ between nodes $v$ and $w$ exists whenever a directed time-stamped edge $(v,w;t)$ exists in the temporal network for at least one time stamp $t$.
Capturing the intensity of interactions, we define \emph{edge weights} in the time-aggregated network as the number of times an edge occurs in the temporal network.
A convenient way to illustrate temporal networks are \emph{time-unfolded representations}.
In this representation, time is unfolded into an additional topological dimension by replacing nodes $v$ and $w$ by temporal copies $v_t$ and $w_t$ for each time step $t$.
Time-stamped edges $(v,w;t)$ are represented by directed edges $(v_t, w_{t+1})$, whose directionality captures the directionality of time.
Finally, we define a \emph{time-respecting path} of length $n$ as a sequence of $n$ time-stamped edges $(v_1,v_2;t_1), (v_2,v_3;t_2), \ldots, (v_{n-1}, v_n;t_n)$ with $t_1 < t_2 < \ldots < t_n$.
In addition, it is common practice to assume a \emph{limited waiting time} $\tau$ for time-respecting paths, additionally imposing the constraint that consecutive interactions occur within a time window of $\tau$, i.e. $0 < t_i-t_{i-1} \leq \tau$ for $i=2, \ldots, n$.
We refer to time-respecting paths of length two as \emph{two-paths}.
Representing the shortest possible time-ordered interaction sequence, two-paths are the simplest possible extension of edges (which can be viewed as ``one-paths'') that capture causality in temporal networks.
As such two-paths are a particularly simple abstraction that allows to study causality in temporal networks~\cite{Pfitzner2013,Rosvall2013}.

\begin{figure}[!hbt]
 \includegraphics[width=\textwidth]{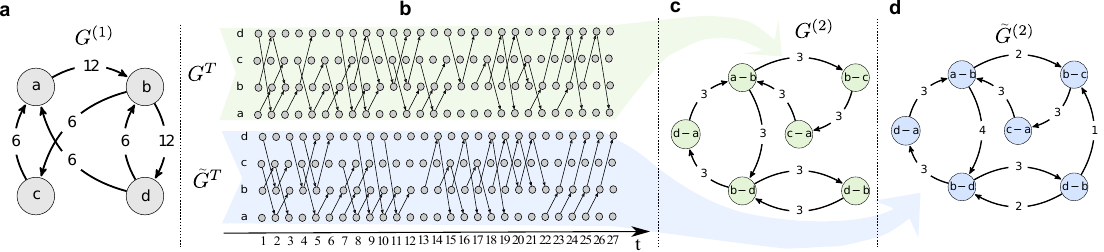}
 \caption{{\bf Two temporal networks with the same first-order, but different second-order time-aggregated networks}
  (a) Time-aggregated network $G^{(1)}$, whose edge weights capture the number of times each edge occurred in a temporal network. The time-aggregated network is consistent with both temporal networks shown in (b).
  (b) Time-unfolded representations of two temporal networks, each consisting of four nodes and $27$ time steps, both consistent with $G^{(1)}$. Differences in their causality structures are highlighted by the corresponding second-order aggregate networks shown in (c) and (d). Both second-order aggregate networks are consistent with $G^{(1)}$.}
 \label{fig:example}
\end{figure}
\begin{figure}[t]
\includegraphics[width=\textwidth]{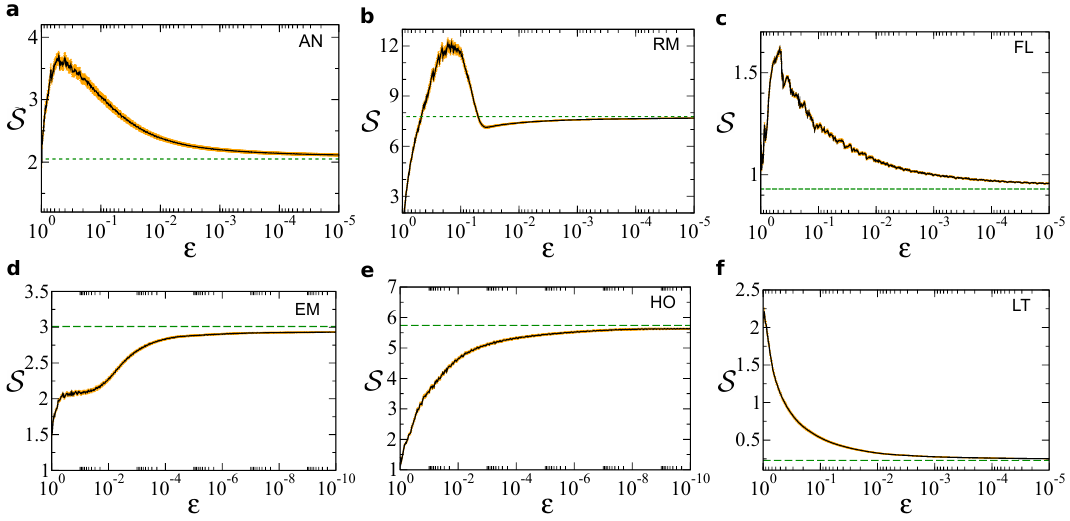}
 \caption{{\bf Causality-driven changes of diffusion speed} We investigate the causality-driven changes of diffusive behaviour by an empirical study of the convergence time of random walks.
For a given convergence threshold $\epsilon$, we compute a slow-down factor $\mathcal{S}(\epsilon)$ which captures the slow-down of diffusion in a temporal network model that preserves the weighted aggregate network as well as the statistics of time-respecting paths of length two in the data, compared to diffusion in the weighted aggregate network.
The six panels show the $\epsilon$-dependent slow-down factor for (a) the (AN) data set covering interactions  between 89 ants, (b) the (RM) data on proximity relations between 64 academic staff members and students, (c) the (FL) data on flight itineraries connecting 116 airports, (d) the (EM) data covering E-Mail exchanges between 167 employees in a company, (e) the (HO) data on contacts between 75 patients and health-care workers in a hospital, and (f) the (LT) data on passenger journeys between 309 London Tube stations. Each result is the mean of random walks starting at every node, error bars indicate the s.e.m. The predicted $\mathcal{S^*}$ value (see Eq.~\ref{eqn:slowdown}) is shown by the horizontal dashed line.}
 \label{fig:tvd}
\end{figure}

Fig.~\ref{fig:example} (b) shows time-unfolded representations of two different temporal networks $G^T$ and $\tilde{G}^{T}$ consisting of four nodes and $27$ time steps.
While both examples correspond to the same weighted time-aggregated network shown in Fig.~\ref{fig:example} (a), the two temporal networks differ in terms of the \emph{ordering} of interactions.
As a consequence, assuming a limited waiting time of $\tau=1$, the time-unfolded representations reveal that a time-respecting path $d \rightarrow b \rightarrow c$ only exists in the temporal network $\tilde{G}^T$, while it is absent in $G^{T}$.
This simple example illustrates how the mere ordering of interactions influences causality in temporal networks.
In the following, we highlight the relevance of causality in real-world systems by studying diffusion dynamics in six empirical temporal network data sets: (AN) time-stamped interactions between ants in a colony~\cite{Blonder2011}; (RM) time-stamped social interactions between students and academic staff at a university campus~\cite{Eagle2006}; (FL) time-ordered flight itineraries connecting airports in the US; (EM) time-stamped E-Mail exchanges between employees of a company~\cite{Michalski2011}; (HO) time-stamped interactions between patients and medical staff in a hospital~\cite{Vanhems2013}; and (LT) passenger itineraries in the London Tube metro system (see details in Methods section).
For each system, we study causality-driven changes of diffusion speed.
In particular, we utilise a random walk process and study the time needed until node visitation probabilities converge to a stationary state~\cite{Lovasz1993,Blanchard2011}.
This convergence behaviour of a random walk is a simple proxy that captures the influence of both the topology and dynamics of temporal networks on general diffusive processes~\cite{Noh2004}.
For a given convergence threshold $\epsilon$, we compute a slow-down factor $\mathcal{S}(\epsilon)$ which captures the slow-down of diffusive behaviour between the weighted aggregated network and a temporal network model derived from the empirical contact sequence respectively (details in Methods section).
In order to exclude effects related to node activities and inter-event time distributions and to exclusively focus on effects of causality observed in the real data sets, this model only preserves the weighted aggregate network as well as the statistics of two-paths in the data.
Fig.~\ref{fig:tvd} shows the causality-driven slow-down factor for the six empirical networks and different convergence thresholds $\epsilon$.
Even though networks are of comparable size, deviations from the corresponding aggregate networks in the limit of small $\epsilon$ (i.e. the long-term behaviour) are markedly different.
For $\epsilon=10^{-5}$ and (RM) the slow-down factor is $\mathcal{S} \approx 7.68\pm0.01$, while for (AN) we obtain a slow-down $\mathcal{S} \approx 2.11\pm0.02$.
For a threshold of $\epsilon=10^{-10}$, in the (HO) data set we have $\mathcal{S} \approx 5.63\pm0.019$, while for (EM) we get $\mathcal{S} \approx 2.93\pm0.005$.
While all these results signify a \emph{slow-down} of diffusion, for $\epsilon=10^{-5}$ and (FL) and (LT) we obtain $\mathcal{S} \approx 0.957\pm0.002$ and $\mathcal{S} \approx 0.25\pm0.001$, which translate to a \emph{speed-up} of diffusion by a factor of $1.04$ and $4$ respectively.
While it is not surprising that the travel patterns in (FL) and (LT) are ``optimised'' in such a way that diffusion is more efficient than in temporal networks generated by contacts between humans (RM, EM and HO) or ants (AN), an analytical explanation for the direction and magnitude of this phenomenon, as well as for the variations across systems, is currently missing.
\subsection{Causality-preserving time-aggregated networks}
In the following we provide an analytical explanation for the direction of this change (i.e. slow-down or speed-up) as well as for its magnitude in specific temporal networks.
We show that an accurate analytical estimate $\mathcal{S^*}$ for the slow-down $\mathcal{S}$ observed in empirical temporal networks can be calculated based on the eigenvalue spectrum of higher-order, time-aggregated representations of temporal networks.
Our approach utilises a \emph{state space expansion} to obtain a higher-order Markovian representation of non-Markovian temporal networks~\cite{Nelson1995}.
This means that a non-Markovian \emph{sequence of interactions} in which the next interaction only depends on the previous one (i.e. one-step memory), can be modeled by a Markovian stochastic process that generates a \emph{sequence of two-paths}.
Analogous to a \emph{first-order time-aggregated network} $G^{(1)}$ consisting of (first-order) nodes $V^{(1)}$ and (first-order) edges $E^{(1)}$, we define a \emph{second-order time-aggregated network} $G^{(2)}$ consisting of \emph{second-order nodes} $V^{(2)}$ and \emph{second-order edges} $E^{(2)}$.
Similar to a directed line graph construction~\cite{Harary1960}, each second-order node represents an edge in the first-order aggregate network.
As second-order edges, we define all possible paths of length two in the first-order aggregate network, i.e. the set of all pairs $\left(e_1, e_2\right)$ for edges $e_1=(a,b)$ and $e_2=(b,c)$ in $G^{(1)}$.
With this, \emph{second-order edge weights} $w^{(2)}(e_1,e_2)$ can be defined as the relative frequency of time-respecting paths $(a,b;t_1) \rightarrow (b,c;t_2)$ of length two in a temporal network.
While the full details of this construction can be found in the Methods section, we illustrate our approach using the two temporal networks shown in Fig.~\ref{fig:example}.
Panels (c) and (d) show two second-order time-aggregated networks $G^{(2)}$ and $\tilde{G}^{(2)}$ corresponding to the temporal networks $G^{T}$ and $\tilde{G}^{T}$ respectively.
In particular, the absence of a time-respecting path $d \rightarrow b \rightarrow c$ in $G^{T}$ is captured by the absence of the second-order edge between the second-order nodes $e_1=\left(d,b\right)$ and $e_2=\left(b,c\right)$.
Further differences between the causality structures of $G^{T}$ and $\tilde{G}^{T}$ are captured by different second-order edge weights.
Notably, this example illustrates that temporal networks giving rise to different second-order time-aggregated networks can still be consistent with the same first-order time-aggregated network.

This approach allows us to generate a second-order network representation where  each second-order node represents an edge in the underlying temporal network, each second-order edge represents a time-respecting path of length two, and weights $w^{(2)}$ capture the statistics of two-paths in the temporal network.
An interesting aspect of this construction is that it allows to easily define \emph{second-order Markov models} generating contact sequences which exhibit ``one-step memory'' and which thus correctly reproduce the statistics of time-respecting paths of length two in the original temporal network.
For this, one can define a \emph{second-order transition matrix} $\mathbf{T}^{(2)}$ where transition rates between second-order nodes are proportional to second-order edge weights (see details in Methods section).
In the following, we illustrate the construction of second-order transition matrices using the examples in Fig.~\ref{fig:example}.
For the second-order aggregate network $G^{(2)}$ shown in panel (c), corresponding to the temporal network $G^T$, the transition matrix $\mathbf{T}^{(2)}$ (rows/columns ordered as indicated) is
\begin{equation*}
    \mathbf{T}^{(2)} =
    \begin{array}{l|}
        (a,b)  \\
        (b,c) \\
        (b,d) \\
        (c,a) \\
        (d,a) \\
        (d,b)
    \end{array}
    \left(
    \begin{array}{cccccc}
      0 & 1/2 & 1/2 & 0 & 0 & 0 \\
      0 & 0 & 0 & 1 & 0 & 0 \\
      0 & 0 & 0 & 0 & 1/2 & 1/2 \\
      1 & 0 & 0 & 0 & 0 & 0 \\
      1 & 0 & 0 & 0 & 0 & 0 \\
      0 & 0 & 1 & 0 & 0 & 0
    \end{array}
    \right).
    \label{eqn:ExampleT2}
\end{equation*}
The leading eigenvector of a stochastic matrix captures the stationary distribution of the associated Markov chain.
As such, the leading eigenvector $\boldsymbol\pi$ of the second-order transition matrix captures the stationary activation frequencies of edges in contact sequences generated by the corresponding second-order Markov model.
For the example above, we obtain a normalised leading eigenvector $\boldsymbol\pi = \left( \frac{1}{4}, \frac{1}{8}, \frac{1}{4}, \frac{1}{8}, \frac{1}{8}, \frac{1}{8} \right)$, which reproduces the relative weights of edges in the first-order aggregate network shown in Fig.~1 (a).
In summary, interpreting $\mathbf{T}^{(2)}$ as transition matrix of a random walker in the second-order aggregate network, we obtain a second-order Markov model generating contact sequences that preserve the relative weights in the first-order aggregate network, as well as the statistics of two-paths.
In line with recent observations that one-step memory is often sufficient to characterise time-respecting paths in empirical temporal networks~\cite{Rosvall2013}, in the remainder of this article we focus on such second-order models.
However, our findings can be generalised to $n$-th order networks $G^{(n)}$ and matrices $\mathbf{T}^{(n)}$ that capture the statistics of time-respecting paths of any length $n$.
From this perspective, the weighted first-order aggregate network can be seen as a first-order approximation where weights only capture the statistics of edges, i.e. time-respecting paths of length one.
Contact sequences generated by a random walk in the first-order time-aggregated network with transition probabilities proportional to edge weights preserve the statistics of edges but destroy the statistics of time-respecting paths.
As such, a random walker in the first-order time-aggregate network must be interpreted as null model that destroys causality, and which can thus not be used to gain analytical insights about dynamical processes in non-Markovian temporal networks~\cite{Barrat2013}.
A second-order representation of the same null model can be constructed using a \emph{maximum entropy second-order transition matrix} $\mathbf{\tilde{T}}^{(2)}$.
For two links $e_1=(a,b)$ and $e_2=(b,c)$, the transition probability $\tilde{T}^{(2)}_{e_1e_2}$ simply corresponds to the transition rate of a random walk across the weighted link $(b,c)$ in the first-order aggregate network (see details in Methods section).
This definition ensures that the corresponding random walker generates Markovian temporal networks which are consistent with a given weighted time-aggregated network, and which exhibit a two-path statistic as expected based on paths in the first-order aggregate network.
We again illustrate our approach using the first-order time-aggregated network $G^{(1)}$ shown in the left panel of Fig.~\ref{fig:example}.
For this example, the transition matrix corresponding to a ``Markovian'' temporal network is given as
\begin{equation*}
    \mathbf{\tilde{T}}^{(2)} =
     \begin{array}{l|}
        (a,b)  \\
        (b,c) \\
        (b,d) \\
        (c,a) \\
        (d,a) \\
        (d,b)
    \end{array}
    \left(
    \begin{array}{cccccc}
      0 & 1/3 & 2/3 & 0 & 0 & 0 \\
      0 & 0 & 0 & 1 & 0 & 0 \\
      0 & 0 & 0 & 0 & 1/2 & 1/2 \\
      1 & 0 & 0 & 0 & 0 & 0 \\
      1 & 0 & 0 & 0 & 0 & 0 \\
      0 & 1/3 & 2/3 & 0 & 0 & 0 \\
    \end{array}
    \right).
\label{eqn:ExampleNullModel}
\end{equation*}
Again, as leading eigenvector we obtain $\boldsymbol\pi = \left( \frac{1}{4}, \frac{1}{8}, \frac{1}{4}, \frac{1}{8}, \frac{1}{8}, \frac{1}{8} \right)$, confirming that the stationary activation frequencies of edges correspond to the relative weights of edges in the first-order time-aggregated network.
From the perspective of statistical ensembles, which is commonly applied in the study of complex networks, each second-order transition matrix whose leading eigenvector $\boldsymbol\pi$ satisfies $\left(\boldsymbol\pi\right)_e = w^{(1)}(a,b)$ ($\forall$ edges $e=(a,b)$) defines a statistical ensemble of temporal networks constrained by a weighted time-aggregated network $G^{(1)}$ and a given two-path statistics.
The entropy $H(\mathbf{T}^{(2)})$ of this ensemble can be defined as the entropy growth rate of the Markov chain described by the corresponding transition matrix (details in Methods section)~\cite{Cover2006}.
Different from entropy measures previously applied to dynamic networks \cite{Zhao2011}, this measure quantifies to what extent the next step in a contact sequence is determined by the previous one.
For a specific second-order transition matrix $\mathbf{T}^{(2)}$ and a corresponding maximum entropy model $\mathbf{\tilde{T}}^{(2)}$, we define the \emph{entropy growth rate ratio} as
\begin{equation}
  \Lambda_H(\mathbf{T}^{(2)}) := H(\mathbf{T}^{(2)})/H(\mathbf{\tilde{T}}^{(2)}).
\label{eqn:entropyDiff}
\end{equation}
This ratio ranges between a minimum of zero for transition matrices corresponding to contact sequences that are completely deterministic, and a maximum of one for transition matrices corresponding to Markovian temporal networks.
In general, an entropy growth rate ratio smaller than one highlights that the statistics of two-paths - and thus causality in the temporal network - deviates from what is expected based on the first-order aggregate network.
As such, $\Lambda_H$ is a simple measure that quantifies the importance of non-Markovian properties in temporal networks.
We illustrate this using the simple example introduced in Fig.~\ref{fig:example}.
For the second-order transition matrices $\mathbf{T}^{(2)}$ and $\mathbf{\tilde{T}}^{(2)}$ we obtain $\Lambda_H(\mathbf{T}^{(2)}) = 0.84$ and thus $\Lambda_H(\mathbf{T}^{(2)})<1$.
This confirms that $\mathbf{T}^{(2)}$ corresponds to a non-Markovian temporal network, and that the statistics of time-respecting paths in $G^{T}$ deviates from what one could expect based on edge frequencies in the first-order aggregate network.
Considering the temporal network $\tilde{G}^{T}$, one easily verifies that edge weights in the corresponding second-order aggregate network $\tilde{G}^{(2)}$ coincide with the transition matrix $\mathbf{\tilde{T}}^{(2)}$.
The resulting entropy growth rate ratio of one for $\tilde{G}^{T}$ verifies that this temporal network does not exhibit non-Markovian characteristics and that two-path statistics do not deviate from what is expected based on the first-order aggregate network.

\subsection{Predicting causality-driven changes of diffusion speed}
A particularly interesting aspect of the second-order network representation introduced above is that \emph{temporal transitivity} is preserved, i.e. the existence of two second-order edges $(e_1,e_2)$ and $(e_2,e_3)$ implies that a time-respecting path $e_1 \rightarrow e_2 \rightarrow e_3$ exists in the underlying temporal network.
Notably, the same is not true for first-order aggregate networks, which do not necessarily preserve temporal transitivity in terms of time-respecting paths; i.e. the existence of two first-order edges $(a,b)$ and $(b,c)$ does not imply that a time-respecting path $a \rightarrow b \rightarrow c$ exists.
Transitivity of paths is a precondition for the use of algebraic methods in the study of dynamical processes.
As such, it is possible to study diffusion dynamics in temporal networks based on the spectral properties of the matrix $\mathbf{T}^{(2)}$, while the same is not true for a transition matrix defined based on edge weights in the first-order aggregate network.
In particular, the convergence time of a random walk process (and thus diffusion speed) can be related to the second largest eigenvalue of its transition matrix~\cite{Chung2005}.
For a primitive stochastic matrix with (not necessarily real) eigenvalues $1=\lambda_1 > |\lambda_2| > |\lambda_3| \geq \ldots \geq |\lambda_n|$, one can show that the number of steps $k$ after which the total variation distance $\Delta(\boldsymbol\pi_k, \boldsymbol\pi)$ between the visitation probabilities $\boldsymbol\pi_k$ and the stationary distribution $\boldsymbol\pi$ of a random walk falls below $\epsilon$ is proportional to $1/\ln(|\lambda_2|)$ (see Supplementary Note 1 for a detailed derivation).
For a matrix $\mathbf{T}^{(2)}$ capturing the statistics of two-paths in an empirical temporal network, and a matrix $\mathbf{\tilde{T}}^{(2)}$ corresponding to the ``Markovian'' null model derived from the first-order aggregate network, an analytical prediction $\mathcal{S}^*$ for causality-driven changes of convergence speed can thus be derived as
\begin{equation}
\mathcal{S}^*(\mathbf{T}^{(2)}) := \ln(|\tilde{\lambda}_2|)/\ln(|\lambda_2|),
\label{eqn:slowdown}
\end{equation}
where $\lambda_2$ and $\tilde{\lambda}_2$ denote the second largest eigenvalue of $\mathbf{T}^{(2)}$ and $\mathbf{\tilde{T}}^{(2)}$ respectively.
Depending on the eigenvalues $\lambda_2$ and $\tilde{\lambda}_2$, both a slow-down ($\mathcal{S}^*(\mathbf{T}^{(2)})>1$) or speed-up ($\mathcal{S}^*(\mathbf{T}^{(2)})<1$) of diffusion can occur.

This approach allows us to analytically study the effect of non-Markovian characteristics in the empirical data sets introduced above.
For each data set we construct matrices $\mathbf{T}^{(2)}$ and $\mathbf{\tilde{T}}^{(2)}$ (see Eqs.~\ref{eqn:T2} and \ref{eqn:T2Null} in Methods), and compute the entropy growth rate ratio $\Lambda_H$ for the corresponding statistical ensembles.
For (RM) we obtain $\Lambda_H(\mathbf{T}^{(2)}) \approx 0.40$, for (AN) $\Lambda_H(\mathbf{T}^{(2)}) \approx 0.42$, for (EM) we get $\Lambda_H(\mathbf{T}^{(2)}) \approx 0.62$ and for (HO) we obtain $\Lambda_H(\mathbf{T}^{(2)}) \approx 0.71$.
For (LT) we obtain $\Lambda_H(\mathbf{T}^{(2)}) \approx 0.30$, while for (FL) we have $\Lambda_H(\mathbf{T}^{(2)}) \approx 0.82$.
This indicates that the topologies of time-respecting paths in all six cases differ from what is expected from the first-order time-aggregated networks.
The impact of these differences on diffusion can be quantified by means of the analytical prediction $\mathcal{S}^*(\mathbf{T}^{(2)})$:
For (RM) we obtain $\mathcal{S}^*(\mathbf{T}^{(2)})\approx 7.77$, for (AN) $\mathcal{S}^*(\mathbf{T}^{(2)}) \approx 2.05$, for (EM) we get $\mathcal{S}^*(\mathbf{T}^{(2)}) \approx 3.01$ and for (HO) we obtain $\mathcal{S}^*(\mathbf{T}^{(2)}) \approx 5.75$.
Considering the two data sets which show a speed-up of diffusion, we get $\mathcal{S}^*(\mathbf{T}^{(2)}) \approx 0.93$ for (FL), while for (LT) we obtain $\mathcal{S}^*(\mathbf{T}^{(2)}) \approx 0.23$.
All six predictions are consistent with the diffusion behaviour observed in numerical simulations in the limit of small $\epsilon$ (see Fig.~\ref{fig:tvd}).
As argued above, the significantly smaller magnitude of the slow-down effect in (AN) compared to (RM) can neither be attributed to differences in system size nor inter-event time distributions.
A spectral analysis of $\mathbf{T}^{(2)}$ can explain the smaller slow-down of (AN) compared to (RM) by a ``better connected'' causal topology indicated by a smaller $\mathcal{S}^*$.
Similarly, the large slow-down observed in (HO) can be related to a ``badly connected'' causal topology indicated by a large value of $\mathcal{S}^*$.
For (FL), the analytical prediction $\mathcal{S}^*(\mathbf{T}^{(2)}) \approx 0.93$ is consistent with the asymptotic empirical speed-up observed in Fig.~\ref{fig:tvd}.
Similarly, the prediction $\mathcal{S}^*(\mathbf{T}^{(2)}) \approx 0.23$ for (LT) is in line with the speed-up observed in Fig.~\ref{fig:tvd}.
Here, the small value of $\mathcal{S}^*(\mathbf{T}^{(2)})$ highlights that the empirical second-order aggregate network is much better connected that one would expect from a Markovian temporal network, thus explaining the large speed-up by a factor of four.
The non-linear behaviour of $\mathcal{S}(\epsilon)$ can be understood by recalling that Eq.~\ref{eqn:slowdown} makes the simplifying assumption that only $\lambda_2$ contributes to the convergence time, which holds in the limit of small $\epsilon$.
As $\epsilon$ increases, an increasing number of eigenvalues and eigenvectors have non-negligible contributions to the empirical slow-down $\mathcal{S}$.
\subsection{Causality structures can slow-down or speed-up diffusion}
Above, we showed that non-Markovian characteristics alter the causal topology of time-varying complex systems, and that the dynamics of diffusion in such systems can be explained by the resulting changes in the eigenvalue spectrum of higher-order aggregate networks, compared to the first-order aggregate network.
We further analytically found that, depending on the system under study, non-Markovian characteristics can both slow-down or speed-up diffusion dynamics.
In the following, we further investigate the mechanism behind the speed-up and slow-down by a model in which order-correlations can mitigate or enforce topology-driven limitations of diffusion speed.
The model generates non-Markovian temporal networks consistent with a uniformly weighted aggregate network with two interconnected communities, each consisting of a random $4$-regular graph with $50$ nodes.
A parameter $\sigma \in (-1,1)$ controls whether time-respecting paths between nodes in \emph{different} communities are - compared to a ``Markovian'' realisation - over- ($\sigma>0$) or under-represented ($\sigma<0$).
The Markovian case coincides with $\sigma=0$.
An important aspect of this model is that realisations generated for any parameter $\sigma$ are consistent with the same weighted aggregate network.
The parameter $\sigma$ exclusively influences the temporal ordering of interactions, but neither their frequency, topology nor their temporal distribution (see Supplementary Note 1 for model details and mathematical proofs).
Fig.~\ref{fig:model} (a) shows the effect of $\sigma$ on the entropy growth rate ratio $\Lambda_H$ (blue, dashed line) and the predicted slow-down $\mathcal{S}^*$ (black, solid line).
\begin{figure}[!htb]
 \centering
 \includegraphics[width=\textwidth]{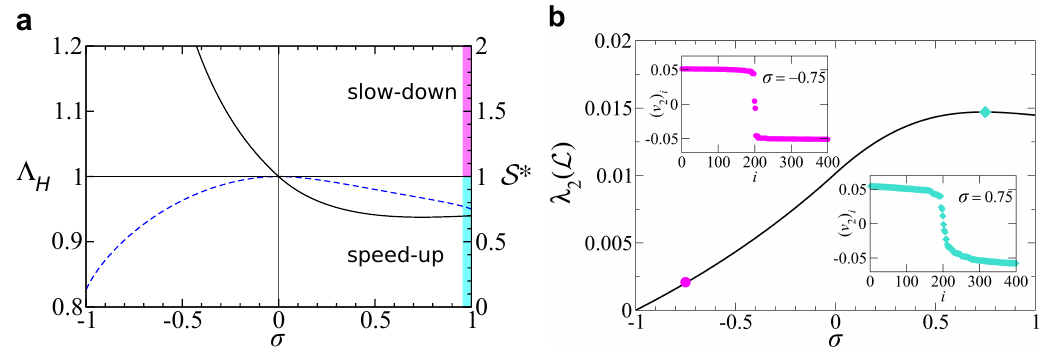}
 \caption{{\bf Slow-down and speed-up regimes in a temporal network model}
 We analytically study a model of non-Markovian temporal networks consistent with a weighted first-order aggregate network with $n=100$ nodes that has two pronounced communities with $50$ nodes each. A parameter $\sigma$ controls whether two-paths across communities are over-represented ($\sigma>0$, turquoise) or under-represented ($\sigma<0$, magenta) compared to the Markovian case ($\sigma=0$). Realisations for different parameters only differ in the ordering of interactions and thus their second-order aggregate networks. Weighted first-order time-aggregated networks are the same for all parameters $\sigma$. (a) Entropy growth rate ratio $\Lambda_H$ (blue, dashed line) and slow-down factor $\mathcal{S}^*$ (black, solid line) for different parameters $\sigma$.
 (b) Algebraic connectivity $\lambda_2(\mathcal{L})$ of the weighted second-order aggregate network for different parameters $\sigma$ of non-Markovian temporal networks. Insets show the Fiedler vector for two points $\sigma=-0.75$ (magenta) and $\sigma=0.75$ (turquoise) in the model's parameter space corresponding to cases where two-paths across communities are inhibited and enforced respectively.}
 \label{fig:model}
\end{figure}
All non-Markovian realisations of the model (i.e. $\sigma \neq 0$) exhibit an entropy growth rate ratio $\Lambda_H<1$ (blue dashed line) which signifies the presence of order correlations.
Whether these correlations result in a speed-up ($\mathcal{S}^*<1$) or slow-down ($\mathcal{S}^*>1$) depends on how order correlations are aligned with community structures.
For $\sigma<0$, time-respecting paths across communities are \emph{inhibited} and diffusion slows down compared to the time-aggregated network ($\mathcal{S}^* > 1$).
For $\sigma>0$, non-Markovian properties \emph{enforce} time-respecting paths across communities and thus \emph{mitigate the decelerating effect of community structures} on diffusion dynamics ($\mathcal{S}^* < 1$)~\cite{Salathe2010}.
We analytically substantiate this intuitive interpretation by means of a a spectral analysis provided in Fig.~\ref{fig:model} (b).
For each $\sigma$, we compute the algebraic connectivity of the causal topology, i.e. the second-smallest eigenvalue $\lambda_2(\mathbf{\mathcal{L}})$ of the normalised Laplacian matrix $\mathbf{\mathcal{L}}=\mathbf{I}_n-\mathbf{T}^{(2)}$ corresponding to the second-order aggregate network ($\mathbf{I}_n$ being the $n$-dimensional identity matrix).
Larger values $\lambda_2(\mathbf{\mathcal{L}})$ indicate ``better-connected'' topologies that do not exhibit \emph{small cuts}~\cite{Fiedler1973,Wu2005}.
The effect of non-Markovian characteristics on $\lambda_2(\mathbf{\mathcal{L}})$ validates that the speed-up and slow-down is due to the ``connectivity'' of the causal topology.
In addition, the insets in Fig.~\ref{fig:model} (b) show entries $(\mathbf{v}_2)_i$ of the Fiedler vector, i.e. the eigenvector $\mathbf{v}_2(\mathcal{L})$ corresponding to eigenvalue $\lambda_2(\mathbf{\mathcal{L}})$.
The distribution of entries of $\mathbf{v}_2(\mathcal{L})$ is related to community structures and is frequently used for divisive spectral partitioning of networks~\cite{Pothen1990}.
For $\sigma=-0.75$, the strong community structure in the causal topology shows up as two separate value ranges with different signs, while the two entries close to zero represent edges that interconnect communities.
Apart from the larger algebraic connectivity, the distribution of entries in the Fiedler vector for $\sigma=0.75$ shows that the separation between communities is less pronounced.
This highlights that non-Markovian properties can effectively outweigh the decelerating effect of community structures in the time-aggregated network, and that the associated changes in the causality structures can be understood by an analysis of the spectrum of higher-order time-aggregated networks.
\section{Discussion}
In summary, we introduce higher-order aggregate representations of temporal networks with non-Markovian contact sequences.
This abstraction allows to define Markov models generating statistical ensembles of temporal networks that preserve the weighted aggregate network as well as the statistics of time-respecting paths.
Focusing on second-order Markov models, we show how transition matrices for such models can be computed based on empirical contact sequences.
The ratio of entropy growth rates (see Eq.~\ref{eqn:entropyDiff}) between this transition matrix and that of a null model, which can easily be constructed from the first-order aggregate network, allows to assess the importance of non-Markovian properties in a particular temporal network.
Considering six different empirical data sets, we show that spectral properties of the transition matrices capture the connectivity of the causal topology of real-world temporal networks.
We demonstrate that this approach allows to analytically predict whether non-Markovian properties slow-down or speed-up diffusive processes as well as the magnitude of this change (see Eq.~\ref{eqn:slowdown}).
With this, we provide the first \emph{analytical explanation} for both the direction and magnitude of causality-driven changes in diffusive dynamics observed in empirical systems.
Focusing on the finding that non-Markovian characteristics of temporal networks can both slow-down or speed-up diffusion processes, we finally introduce a simple model that allows to analytically investigate the underlying mechanisms.
Our results show that the mere ordering of interactions can either mitigate or enforce topological properties that limit diffusion speed.
Both our empirical and analytical studies confirm that causality structures in real-world systems have large and significant effects, slowing down diffusion by a factor of more than seven in one system, while other systems experience a speed-up by a factor of four compared to what is expected from the first-order time-aggregated network.
These findings highlight that the causal topologies of time-varying complex systems constitute an important additional temporal dimension of complexity, which can reinforce, mitigate and even outweigh effects that are due to topological features like, e.g., community structures.

Different from studies exclusively considering how interactions are distributed in time, in our study we focus on how their ordering influences causality structures in temporal networks.
The finding that causality structures alone can lead to both a speed-up or a slow-down of diffusion highlights that, in order to understand the influence of the temporal dynamics in real-world systems, effects of \emph{both} activity patterns and causality must be taken into account.
Considering temporal networks in which interactions are homogeneously distributed in time, our approach further provides a novel explanation for changes in dynamical processes that cannot be explained in terms of bursty node activity patterns.
An additional benefit of our approach is that it can be used for the network-based study of systems for which causal relations between different links can be inferred even though links cannot be assigned absolute time stamps.
The data on airline and subway passenger itineraries analysed in our article are two examples for such systems where only the ordering of links is known.

Our approach of constructing higher-order Markov models that preserve the statistics of time-respecting paths allows to study the \emph{temporal-topological} dimension of time-varying complex systems - a dimension that is often ignored when exclusively focusing on changes in the \emph{duration} of dynamical processes.
The higher-order time-aggregated networks introduced in this article are simple static representations of temporal networks which - compared to first-order aggregate networks - better preserve causality.
This approach provides interesting perspectives not only for analytical studies of further classes of dynamical processes in complex systems with time-varying interaction topologies.
It is also a promising approach for the development of novel \emph{temporal community detection algorithms} using, e.g., spectral clustering or random walk based methods as well as for the design of refined eigenvector-based centrality measures taking into account the ordering of links in dynamic networks.
Finally, we foresee applications in the development of novel temporal network visualisation methods, such as layout algorithms that make use of both the first- and the second-order time-aggregated networks.

\section*{Methods}

\subsection*{Details on empirical data sets}
In our article, we study diffusion dynamics in temporal networks constructed from six different empirical data sets: (AN) captures pairwise interactions between individuals in an ant colony, (RM) is based on contact networks of students and academic staff members at a university campus, (EM) covers E-Mail exchanges between employees of a company, (FL) represents multi-segment itineraries of airline passengers in the United States, and (LT) captures passenger journeys in the London underground transportation network.

For the (AN) data set, we used the largest data set from an empirical study of ant interactions~\cite{Blonder2011}, i.e. the first filming of colony $1$ with a total of $1911$ antenna-body interactions between $89$ ants recorded over a period of $1438$ seconds.
For the (RM) data set, we used time-stamped proximity data on students and academic staff members recorded via Bluetooth-enabled phones at a university campus over a period of more than six months~\cite{Eagle2006}.
For computational reasons, we used a subset covering the week from Sept. 8th to 15th 2004, which comprises a total of $26,260$ time-stamped interactions between $64$ individuals.
The (EM) data set covers E-Mail exchanges recorded over a period of nine months between $167$ employees of a medium-size manufacturing company~\cite{Michalski2011}.
Here, we use a subset covering close to $11,000$ E-mail exchanges occurring during the first month of the observation period.
The (HO) data set contains time-stamped contacts between $46$ health-care workers and $29$ patients in a hospital in Lyon~\cite{Vanhems2013}.
Contacts have been recorded via proximity sensing badges in the week from Dec. 6 to Dec. 10 2010.
For our analysis, we use a subset of more than $15,000$ contacts occurring within the first $48$ hours of the observation period.
The (FL) data set has been extracted from the freely available RITA TranStats Airline Origin and Destination Survey (DB1B) database~\cite{FLdata}, which contains 10 \% samples of all airline tickets sold in the United States for each quarter since 1993.
For our study we extracted $~230,000$ multi-segment flights ticketed by American Airlines in the fourth quarter of 2001, which connect a total of $116$ airports in the United States.
For each ticket number $i$, an itinerary consists of a time-ordered sequence of multiple flight segments between airports indicated by their three-letter IATA code.
An example for a time-ordered itinerary with ticket number $i$ is given in the following:
\begin{align*}
i, CLT, ORF \\
i, ORF, LGA \\
i, LGA, ORF \\
i, ORF, CLT
\end{align*}
While no precise time stamps are known for individual segments, their ordering allows to directly construct time-respecting flight paths taken by individual passengers.
For the example above, a time-respecting path $(CLT, ORF; 1) \rightarrow (ORF,LGA; 2)  \rightarrow (LGA, ORF; 3) \rightarrow (ORF,CLT; 4)$ can be constructed.
Here time-respecting paths necessarily consist of interactions which \emph{immediately follow each other in subsequent time steps}, since otherwise a spurious flight path $(CLT, ORF; 1) \rightarrow (ORF,CLT; 4)$ would be inferred for the example above.
Furthermore, a time-respecting path is only inferred if the ticket number of consecutive flight segments is identical.
We used the same approach in the (LT) data set, which has been extracted from the freely available Rolling Origin \& Destination Survey (RODS) database~\cite{LTdata} provided by the London Underground.
The RODS database covers a 5 \% sample of all journeys made by passengers who used the \emph{Oyster} electronic ticketing card during a period of one week.
This amounts to a total of more than four million passenger flows between $309$ London Underground stations.
By mapping those passenger flows to a network representation of the London Underground, we extracted detailed itinerary data just like those in the example for the (FL) data set.
We then computed time-respecting paths based on directly consecutive travel segments in the same way as for the (FL) data set.
While the condition of directly consecutive travel segments is crucial for the (FL) and the (LT) data set, for the (AN), (RM), (HO) and (EM) data sets we relax this definition of a time-respecting path and additionally consider time-respecting paths if links occur within a certain time period.
In particular, following arguments that many dynamical processes set limitations on how long paths are allowed to wait at certain nodes~\cite{Holme2012}, we limit the \emph{waiting time} on time-respecting paths to a maximum of $\tau$.
In other words, we assume that a time-respecting path between nodes $a$ and $c$ exists whenever two time-stamped edges $(a,b;t_1)$ and $(b,c;t_2)$ exist for $0 < t_2-t_1 \leq \tau$.
In general, we have chosen the maximum waiting time  $\tau$ as the smallest possible value such that the set of nodes that can mutually influence each other via time-respecting paths (i.e. the strongly connected component) represents a sizeable fraction of the network.
For the (AN) data, a maximum waiting time $\tau$ of six seconds was applied, which gives rise to a subset of $61$ nodes that can reach each other via time-respecting paths.
For the (RM) and (HO) data sets, we used a maximum waiting time $\tau$ of five minutes, which resulted in a subset of $58$ and $53$ individuals respectively who can mutually reach each other via time-respecting paths.
For the (EM) data set, we used a maximum waiting time $\tau$ of $60$ minutes, resulting in a subset of $96$ employees mutually connected via time-respecting paths.
For the (FL) data set, the strongly connected component comprises $116$ airports, while it comprises $132$ underground stations in the (LT) data set.

\subsection*{Diffusion dynamics in empirical temporal networks}

We study causality-driven changes of diffusive behaviour in the six temporal network data sets (AN), (RM), (FL), (EM), (HO), (FL) and (LT) described above.
We use the convergence behaviour of a random walk process as a proxy that captures the influence of both the topology and dynamics of temporal networks on general diffusive processes.
For this, we first consider a random walk process in the weighted, time-aggregated network and study the time needed until node visitation probabilities converge to a stationary state.
Starting from a randomly chosen node, in each step of the random walk the next step is chosen with probabilities proportional to the weights of incident edges.
A standard approach to assess the convergence time of random walks is to study the evolution of the \emph{total variation distance} between observed node visitation probabilities and the stationary distribution~\cite{Rosenthal1995}.
For a distribution $\boldsymbol\pi_k$ of visitation probabilities $\left(\boldsymbol\pi_k\right)_v$ of nodes $v$ after $k$ steps of a random walk and a stationary distribution $\boldsymbol\pi$, the total variation distance is defined as
\begin{align*}
  \Delta(\boldsymbol\pi_k,\boldsymbol\pi):=\frac{1}{2}\sum_v{|\left(\boldsymbol\pi\right)_v-\left(\boldsymbol\pi_k\right)_v|}.
\end{align*}
For a given threshold distance $\epsilon$, we define the convergence time $t_{agg}(\epsilon)$ as the minimum number of steps $k$ after which $\Delta(\boldsymbol\pi_k,\boldsymbol\pi) < \epsilon$.
The random walk itineraries produced by this simple random walk model correctly reproduce edge weights in the time-aggregated network and the use of random walk itineraries as a model for temporal networks has been proposed before~\cite{Barrat2013}.
However, random walk itineraries do not preserve statistics of longer time-respecting paths and thus alter causality.
In order to derive a causality-driven slow-down factor, we thus contrast the convergence time $t_{\text{agg}}(\epsilon)$ with the convergence time $t_{\text{temp}}(\epsilon)$ of a second model that additionally preserves the statistics of time-respecting paths of length two in the real data sets (see previous section for details on how we define time-respecting paths in the different data sets).
Again starting with a random node, this model randomly chooses two-paths according to their relative frequencies in the data set, thus corresponding to a walk process which is advanced by two steps at a time.
The random itineraries generated by this model correctly reproduce edge weights in the time-aggregated network, and - different from a random walk in the time-aggregated network - the statistics of time-respecting paths of length two.
For a given threshold distance $\epsilon$, we again define the convergence time $t_{\text{temp}}(\epsilon)$ as the minimum number of steps $k$ after which $\Delta(\boldsymbol\pi_k,\boldsymbol\pi) < \epsilon$.
For a convergence threshold $\epsilon$, this allows us to define a causality-driven slow-down factor $\mathcal{S}(\epsilon) := t_{\text{temp}}(\epsilon)/t_{\text{agg}}(\epsilon)$ that is due to the \emph{temporal-topological} characteristics of time-respecting paths, while ruling out effects of inter-event time distributions or bursty node activities.

\subsection*{Constructing higher-order time-aggregated networks}

Extracting time-respecting paths in the six data sets allows us to construct higher-order time-aggregated representations of the underlying temporal interaction sequences.
In the following, we provide a detailed description of this construction.
We consider a temporal network $G^{T}$ consisting of directed time-stamped edges $(v,w;t)$ for nodes $v$ and $w$ and discrete time stamps $t$.
A \emph{first-order time-aggregated network} $G^{(1)}$ can then be defined, where a directed edge $(v,w)$ between nodes $v$ and $w$ exists whenever a time-stamped edge $(v,w;t)$ exists in $G^T$ for some time stamp $t$.
In addition, edge weights $w^{(1)}\left(v,w\right)$ can be defined as the (relative) number of edge occurrences in the temporal network.
Considering that edges can be thought of as time-respecting paths of length one, we can similarly construct a \emph{second-order time-aggregated network} by considering time-respecting paths of length two.
For this, we define a second-order time-aggregated network $G^{(2)}$ as tuple $(V^{(2)}, E^{(2)})$ consisting of \emph{second-order nodes} $V^{(2)}$ and \emph{second-order edges} $E^{(2)}$.
Second-order nodes $e \in V^{(2)}$ represent edges in the first-order aggregate network $G^{(1)}$.
Second-order edges $E^{(2)}$ represent all possible time-respecting paths of length two in $G^{(1)}$.
Based on the definition of time-respecting paths with a limited waiting time $\tau$, \emph{second-order edge weights} $w^{(2)}(e_1,e_2)$ can be defined based on the frequency of two-paths, i.e. the frequency of time-respecting paths $(a,b;t_1) \rightarrow (b,c;t_2)$ of length two in $G^{T}$ (for $1 \leq t_2-t_1 \leq \tau$).
Since multiple two-paths $(a',b;t) \rightarrow (b,c';t')$ can pass through node $b$ at the same time, it is necessary to proportionally correct second-order edge weights for all multiple occurrences.
For the simple case $\tau=1$, one can define second-order edge weights as
\begin{equation}
  w^{(2)}\left(e_1,e_2\right) := \sum_{t}{ \frac{\delta_{(a,b;t-1)} \delta_{(b,c;t)}}{\sum_{a', c' \in V}{\delta_{(a',b;t-1)}\delta_{(b,c';t)}}}},
\end{equation}
where $\delta_{(a,b;t)}=1$ if edge $(a,b;t)$ exists in the temporal network $G^{T}$ and $\delta_{(a,b;t)}=0$ otherwise.
Following the arguments above, it is simple to generalise weights to capture two-paths $(a,b;t_1)-(b,c;t_2)$ for $1 \leq t_2-t_1 \leq \tau$.
The software used to infer time-respecting paths and to construct weighted second-order time-aggregated networks from the six empirical data sets is available online~\cite{code}.

\subsection*{Higher-order Markov models for temporal networks}

Using the second-order time-aggregated network $G^{(2)}$ and second-order edge weights $w^{(2)}$ defined above, for all time-respecting paths $e_1 \rightarrow e_2$ of length two we define the entries of the transition matrix $\mathbf{T}^{(2)}$ for a random walk in the weighted network $G^{(2)}$ as
\begin{equation}
  T^{(2)}_{e_1e_2} := w^{(2)}\left(e_1,e_2\right) \left( \sum_{e' \in V^{(2)}}{w^{(2)}\left(e_1,e'\right)} \right)^{-1}.
  \label{eqn:T2}
\end{equation}
In line with the standard way of defining random walks on weighted networks, transition rates between nodes $e_1$ and $e_2$ are defined to be proportional to edge weights and are normalised by the cumulative weight of all edges $(e_1, e')$ emanating from node $e_1$.
If the transition matrix $\mathbf{T}^{(2)}$ is primitive, the Perron-Frobenius theorem guarantees that a unique leading eigenvector $\boldsymbol\pi$ of $\mathbf{T}^{(2)}$ exists.
Note that $\mathbf{T}^{(2)}$ can always be made primitive by restricting it to the largest strongly connected component of $G^{(2)}$ and adding small positive diagonal entries.

While the transition matrix $\mathbf{T}^{(2)}$ captures the statistics of two-paths in a given temporal network, we can additionally define a maximum entropy transition matrix $\mathbf{\tilde{T}}^{(2)}$ which captures the two-path statistics one would expect based on the relative edge weights in the first-order time-aggregated network.
For $e_1=(a,b)$ and $e_2=(b,c)$, the entries $\tilde{T}^{(2)}_{e_1e_2}$ corresponding to a two-path $e_1 \rightarrow e_2$ are given as
\begin{equation}
  \tilde{T}^{(2)}_{e_1e_2} := w^{(1)}\left(b,c\right) \left( \sum_{c' \in V^{(1)}}{w^{(1)}\left(b,c'\right)} \right)^{-1}.
\label{eqn:T2Null}
\end{equation}

This second-order Markov model preserves the weights $w^{(1)}$ of edges in $G^{(1)}$ and creates ``Markovian'' temporal networks in which consecutive links are independent from each other.

The entropy of a second-order Markov model for a particular temporal network can be quantified in terms of the entropy growth rate of a transition matrix $\mathbf{T}^{(2)}$.
This notion of entropy quantifies the amount of information that is lost about the current state of a Markov process based on a given transition matrix.
We define the entropy growth rate of a second-order transition matrix as
\begin{equation}
  H(\mathbf{T}^{(2)}) := - \sum_{e \in E^{(1)}} \left(\boldsymbol\pi\right)_e \sum_{e'\in E^{(1)}} T^{(2)}_{ee'} \log_2\left(T^{(2)}_{ee'}\right).
\label{eqn:entropy}
\end{equation}
For a transition matrix which only consists of deterministic transitions with probability $1$, the entropy growth rate is zero, while it reaches a (size-dependent) maximum for a transition matrix where every state can be reached with equal probability in every step.

\subsection*{Software}
We finally remark that our results from above can be reproduced by means of the \textsc{python} package \textsc{pyTempNets}, which is freely available from \url{https://github.com/IngoScholtes/pyTempNets}.

\section*{Acknowledgements}
I.S. acknowledges financial support by SNF project CR\_31I1\_140644.
I.S. and R.P. acknowledge support by the COST action TD1210 KNOWeSCAPE.
N.W., A.G. and F.S. acknowledge financial support by EU-FET project MULTIPLEX 317532.
The authors acknowledge feedback on the manuscript by R. Burkholz.

\section*{Author contributions}
I.S., N.W., R.P., A.G., C.J.T. and F.S. conceived and designed the research.
I.S. and N.W. analysed data, performed the simulations, provided the analytical results and wrote the article.
All authors discussed the results, reviewed and edited the manuscript.

\newpage

\section*{Supplementary Information}

\renewcommand{\thefigure}{\arabic{figure}}
\renewcommand{\theequation}{\arabic{equation}}

\renewcommand{\figurename}{Supplementary Figure}

This supplementary information contains technical details about the derivation of the slow-down factor $\mathcal{S}^*$ as well as details about a model for non-Markovian temporal networks that can be parameterised to produce temporal networks that slow-down or speed-up diffusion.

\subsection*{Derivation of Slow-Down Factor}
In our article, we argue that changes of diffusion dynamics in temporal networks as compared to their static counterparts, are due to the change of \emph{connectedness}, or \emph{conductance}, of the corresponding \emph{second-order aggregate network}.
We further show that these changes are captured by a slow-down factor which can be computed based on the second-order aggregate networks corresponding to a particular non-Markovian temporal network and its Markovian counterpart.
In the following, we substantiate our approach by analytical arguments, highlighting the conditions under which our prediction is accurate.

For a second-order aggregate network $G^{(2)}$ with a weight function $w^{(2)}$, let us consider a transition matrix $\mathbf{T}^{(2)}$ as defined in Eq.~2 of our article.
The influence of the eigenvalues of $\mathbf{T}^{(2)}$ on the convergence behavior of a random walk can then be studied as follows.
For a sequence of eigenvalues $1=\lambda_1 \geq |\lambda_2| \geq \ldots \geq |\lambda_n|$ of $\mathbf{T}^{(2)}$  with corresponding eigenvectors $\mathbf{v}_1, \ldots, \mathbf{v}_n$, we define the \emph{eigenmatrix}  $\mathbf{U} := (\mathbf{v}_i)_{i=1, \ldots ,n}$.
We further define a stochastic row vector $\mathbf{x}=\boldsymbol\pi_0=(p_1, \ldots, p_n)$ which we assume contains the initial node visitation probabilities before the random walk starts.
Since $\mathbf{U}$ is not necessarily regular (n.b. that $G^{(2)}$ is directed) we use a Moore-Penrose pseudoinverse~\cite{Penrose1955} $\mathbf{U}^{-1}$ of $\mathbf{U}$ as well as diagonal matrix $\mathbf{D}=\text{diag}(\lambda_1, \ldots, \lambda_n)$ to obtain an eigendecomposition of $\mathbf{T}^{(2)}$ as
\begin{equation}
\mathbf{T}^{(2)}=\mathbf{U}^{-1}\mathbf{D} \mathbf{U}.
\end{equation}
We can then transform the vector $\mathbf{x}$ into an eigenspace representation of $\mathbf{T}^{(2)}$ and obtain $\mathbf{a}=\mathbf{x}\mathbf{U}^{-1}$ such that $\mathbf{x}=\sum_{i=1}^n{a_i\mathbf{v}_i}$.
With this, the node visitation probability vector $\boldsymbol\pi_k$ after $k$ steps can be expressed as
\begin{equation*}
\boldsymbol\pi_k = \mathbf{x} \mathbf{T}^k = \sum_{i=1}^n{a_i\mathbf{v}_i\mathbf{T}^k}
\end{equation*}
where $\mathbf{T}^k$ is the $k$-th power of the transition matrix $\mathbf{T}$ and $a_i$ is the $i$-th entry of vector $\mathbf{a}$.
Repeated substitution according to the eigenvalue equation $\mathbf{v}_i \mathbf{T} = \lambda_i \mathbf{v}_i$ yields
\begin{equation*}
\boldsymbol\pi_k = \sum_{i=1}^n\lambda_i^k a_i \mathbf{v}_i.
\end{equation*}
Assuming that $\mathbf{T}^{(2)}$ is primitive, for the Perron-Frobenius eigenvalue $\lambda_1$ we obtain $1 = \lambda_1 > |\lambda_2|$ and the normalised first eigenvector $a_1\mathbf{v}_1$ corresponds to the unique stationary distribution $\boldsymbol\pi=\boldsymbol\pi_k$ of the Markov chain given by $\mathbf{T}^{(2)}$.
For the first term in the sum above, we thus obtain $\lambda_1^k a_1 \mathbf{v}_1 = 1 \cdot \boldsymbol\pi = \boldsymbol\pi$.
With
\begin{equation}
\boldsymbol\pi_k = \boldsymbol\pi + \sum_{i=2}^n\lambda_i^k a_i \mathbf{v}_i
\end{equation}
a difference vector $\boldsymbol\delta(k)$ whose components $\delta_j(k)$ capture the difference between node visitation probabilities $\left(\boldsymbol\pi_k\right)_j$ after $k$ steps of the random walk and the stationary visitation probability $\left(\boldsymbol\pi\right)_j$ for each node $j$ can be defined as
\begin{equation}
\boldsymbol\delta(k) = \boldsymbol\pi_k - \boldsymbol\pi = \sum_{i=2}^n\lambda_i^ka_i\mathbf{v}_i.
\end{equation}
The total variation distance
\begin{equation*}
\Delta(\boldsymbol\pi_k,\boldsymbol\pi):=\frac{1}{2}\sum_{j=1}^n|\left(\boldsymbol\pi\right)_j-\left(\boldsymbol\pi_k\right)_j|
\end{equation*}
after $k$ steps can then be given as
\begin{align*}
    \Delta(\boldsymbol\pi_k,\boldsymbol\pi)   & =     & \frac{1}{2} \sum_{j=1}^n |    & \delta_j(k)| \\
                        & =     & \frac{1}{2} \sum_{j=1}^n |    & \lambda_2^k a_2\left(\mathbf{v}_2\right)_j + \lambda_3^k a_3\left(\mathbf{v}_3\right)_j  \\
                        &       &                               & + \ldots + \lambda_n^k a_n\left(\mathbf{v}_n\right)_j |
\end{align*}
where $\left(\mathbf{v}_i\right)_j$ denotes the $j$-th component of the $i$-th eigenvector $\mathbf{v}_i$.
Under the condition that $|\lambda_2|$ is not degenerate (i.e. $|\lambda_2| > |\lambda_3|$) and using the fact that $|\lambda_i|<1$ for $i \geq 2$ (n.b. that $\mathbf{T}^{(2)}$ is primitive and thus $G^{(2)}$ is necessarily strongly connected) for $k$ sufficiently large one can make the following approximation:
\begin{equation*}
\Delta(\boldsymbol\pi_k,\boldsymbol\pi) \approx \frac{1}{2} \sum_{j=1}^n  |\lambda_2^ka_2\left(\mathbf{v}_2\right)_j|.
\end{equation*}
For a sufficiently small convergence threshold $\epsilon>0$, the convergence time $k$ after which the total variation distance falls below $\epsilon$ can then be calculated as follows:
\begin{align*}
\Delta(\boldsymbol\pi_k,\boldsymbol\pi) \approx \frac{1}{2} \sum_{j=1}^n |\lambda_2^ka_2\left(\mathbf{v}_2\right)_j| \leq \epsilon \Leftrightarrow \\
		k \cdot \ln(|\lambda_2|) + \ln\left(\frac{1}{2} \sum_{j=1}^n | a_2\left(\mathbf{v}_2\right)_j|\right) \leq \ln(\epsilon) \Leftrightarrow \\
        k \geq \frac{1}{{\ln(|\lambda_2|)}} \cdot \left( \ln(\epsilon) - \ln\left(\frac{1}{2}\sum_{j=1}^n | a_2 \left(\mathbf{v}_2\right)_j|\right)\right) \\
\end{align*}
Here, we utilise the fact that, since $|\lambda_2| > |\lambda_3|$, both $\lambda_2$ and $a_2\mathbf{v}_2$ are necessarily real and thus $|\lambda_2^ka_2\left(\mathbf{v}_2\right)_j|=|\lambda_2^k|\cdot |a_2\left(\mathbf{v}_2\right)_j|=|\lambda_2|^k\cdot |a_2\left(\mathbf{v}_2\right)_j|$.
Based on the result above, the convergence time $t(\epsilon)$ after which total variation falls below $\epsilon$ (i.e. $\forall k \geq t(\epsilon): \Delta(\boldsymbol\pi_k,\boldsymbol\pi)\leq \epsilon$) is than given as
\begin{equation*}
        t(\epsilon) = \frac{1}{{\ln(|\lambda_2|)}} \cdot \left( \ln(\epsilon) - \ln\left(\frac{1}{2}\sum_{j=1}^n | a_2 \left(\mathbf{v}_2\right)_j|\right)\right).
\end{equation*}
We now consider the null model $\mathbf{\tilde{T}}^{(2)}$ corresponding to a Markovian temporal network model derived from $G^{(2)}$ (and thus to a random walk running on the weighted aggregate network) according to Eq.~3 in our main article.
Based on the sequence of eigenvalues $1=\tilde{\lambda}_1 \geq |\tilde{\lambda}_2| \geq \ldots \geq |\tilde{\lambda}_n|$ of $\mathbf{\tilde{T}}^{(2)}$ with corresponding eigenvectors $\mathbf{\tilde{v}}_1, \ldots, \mathbf{\tilde{v}}_n$, a convergence time $\tilde{t}(\epsilon)$ after which total variation distance falls below $\epsilon$ can then be derived analogously as:
\begin{equation*}
\tilde{t}(\epsilon) = \frac{1}{{\ln(|\tilde{\lambda}_2|)}} \cdot \left(\ln(\epsilon) - \ln\left(\frac{1}{2}\sum_{j=1}^n | \tilde{a}_2\left(\mathbf{\tilde{v}}_2\right)_j|\right)\right)
\end{equation*}
A fraction $\mathcal{S}^*(\mathbf{T}^{(2)}, \epsilon)$ that captures the slow-down (or speed-up) of convergence that is due to non-Markovian properties can then be defined based on $t(\epsilon)/\tilde{t}(\epsilon)$:
\begin{equation*}
\mathcal{S}^*(\mathbf{T}^{(2)}, \epsilon) := \frac{{\ln(|\tilde{\lambda}_2|)}}{{\ln(|\lambda_2|)}} \cdot \frac{\ln(\epsilon)-\ln\left(\frac{1}{2}\sum_{j=1}^n | a_2\left(\mathbf{v}_2\right)_j|\right)}{\ln(\epsilon)-\ln\left(\frac{1}{2}\sum_{j=1}^n | \tilde{a}_2\left(\mathbf{\tilde{v}}_2\right)_j|\right)}
\end{equation*}
We then define the proportional slow-down $\mathcal{S}^*(\mathbf{T}^{(2)})$ in the limit of small $\epsilon$ (or large $k$) as
\begin{equation}
\mathcal{S}^*(\mathbf{T}^{(2)}) := \lim_{\epsilon \rightarrow 0}\left( \mathcal{S}^*(\mathbf{T}^{(2)}, \epsilon) \right) = \frac{{\ln(|\tilde{\lambda}_2|)}}{{\ln(|\lambda_2|)}}.
\end{equation}
We remark, that this slow-down is due to the difference in the spectral gap $1-|\lambda_2|$ of $\mathbf{T}^{(2)}$ as compared to the null-model $\mathbf{\tilde{T}}^{(2)}$ derived from the weighted aggregate network corresponding to both $\mathbf{T}^{(2)}$ and $\mathbf{\tilde{T}}^{(2)}$.
The prediction $\mathcal{S}^*(\mathbf{T}^{(2)})$ holds for sufficiently large $k$ or - equivalently - for a sufficiently small total variation distance $\epsilon$.
Furthermore, we assumed that $\mathbf{\tilde{T}}^{(2)}$ is primitive and that $\lambda_2$ is non-degenerate.

If the gap $1-|\tilde{\lambda}_2|$ of the second-order network corresponding to the Markovian temporal network is larger than the gap $1-|\lambda_2|$ corresponding to a non-Markovian case, $\mathcal{S}^*(\mathbf{T}^{(2)})>1$.
In this case, the \emph{conductance} of $\tilde{G}^{(2)}$ is larger than that of $G^{(2)}$ and the non-Markovian properties slow down random walk convergence.
If - on the other hand - the gap $1-|\tilde{\lambda}_2|$ is smaller than the gap $1-|\lambda_2|$, the conductance of $\tilde{G}^{(2)}$ is smaller than that of $G^{(2)}$.
In this case $\mathcal{S}^*(\mathbf{T}^{(2)})<1$, meaning that the non-Markovian properties of a temporal network speed up random walk convergence.

We finally note that for $|\lambda_2| = |\lambda_3|$, a similar slow-down ratio can be derived for the chi-square distance based on the upper bounds on the second-largest eigenvalues for general directed networks with arbitrary eigenvalue spectra following the arguments put forth in~\cite{Chung2005}.
Based on this approach the prediction would look like
\begin{equation*}
  S^*_{\chi}(\mathbf{T}^{(2)})=\frac{\ln\left(\frac{1}{2}(1+\mathrm{Re}(\tilde{\lambda}_2))\right)}{\ln\left(\frac{1}{2}(1+\mathrm{Re}(\lambda_2))\right)}\, ,
\end{equation*}
with the eigenvalue sequence of the transition matrix sorted by their real parts, i.e. $\mathrm{Re}(\lambda_1) \geq \mathrm{Re}(\lambda_2) \geq \ldots \geq  \mathrm{Re}(\lambda_n)$.
The prediction $S^*_{\chi}(\mathbf{T}^{(2)})$ is equal to $S^*(\frac{1}{2}(\mathbf{I}_n+\mathbf{T}^{(2)}))$ where $n$ is the dimension of $\mathbf{T}^{(2)}$ and $\mathbf{I}_n$ is the corresponding identity matrix.
This is equal to applying the prediction $S^*$ to a transition matrix of a lazy random walk with self-loop probability $1/2$.
This approach can alleviate periodicity and assure that $|\lambda_2| > |\lambda_3|$ at least for the transition matrix of a lazy random walk.

\subsection*{Details of Model for non-Markovian Temporal Networks}

A particularly important finding in our article is the fact that non-Markovian characteristics can give rise both to a slow-down and speed-up of diffusion dynamics when compared to their static aggregated counterparts.
To illustrate this fact, we introduce a simple toy model for temporal networks in which non-Markovian properties can either \emph{inhibit} or \emph{enforce} time-respecting paths across two pronounced communities that are present in the static aggregate network.
In our article we argue that the presence of order correlations which enforce time-respecting paths across communities is a particularly simple mechanism by which non-Markovian properties in temporal networks can speed up diffusion dynamics.
With this we further highlight one possible mechanism by which non-Markovian properties can effectively mitigate the decelerating effect of community structures on diffusion dynamics.

In the following, we formally define our toy model and substantiate our interpretations in the article by means of a spectral analysis of the second-order aggregate networks corresponding to different points in the model's parameter space.
The model is based on a directed, weighted aggregate network $G^{(1)}$ with two communities, each consisting of a random $k$-regular graph with $n$ nodes.
To interconnect the two communities, we randomly draw edges $e=(v_1,v_2)$ and $e'=(v'_1, v'_2)$ from the two communities respectively, remove $e$ and $e'$ and instead add edges $(v_1, v'_1)$ and $(v_2,v'_2)$ thus maintaining a $k$-regular aggregate network.
We further assign uniform weights $\omega_1$ to all edges, thus obtaining a network as shown in the schematic illustration in panel (a) of Fig.~\ref{fig:model}.
For the simulations in the article, we use $k=4$ and $n=50$, thus obtaining a network with $100$ nodes and $400$ directed edges.

\begin{figure}[!ht]
 \centering
 \includegraphics[width=.5\textwidth]{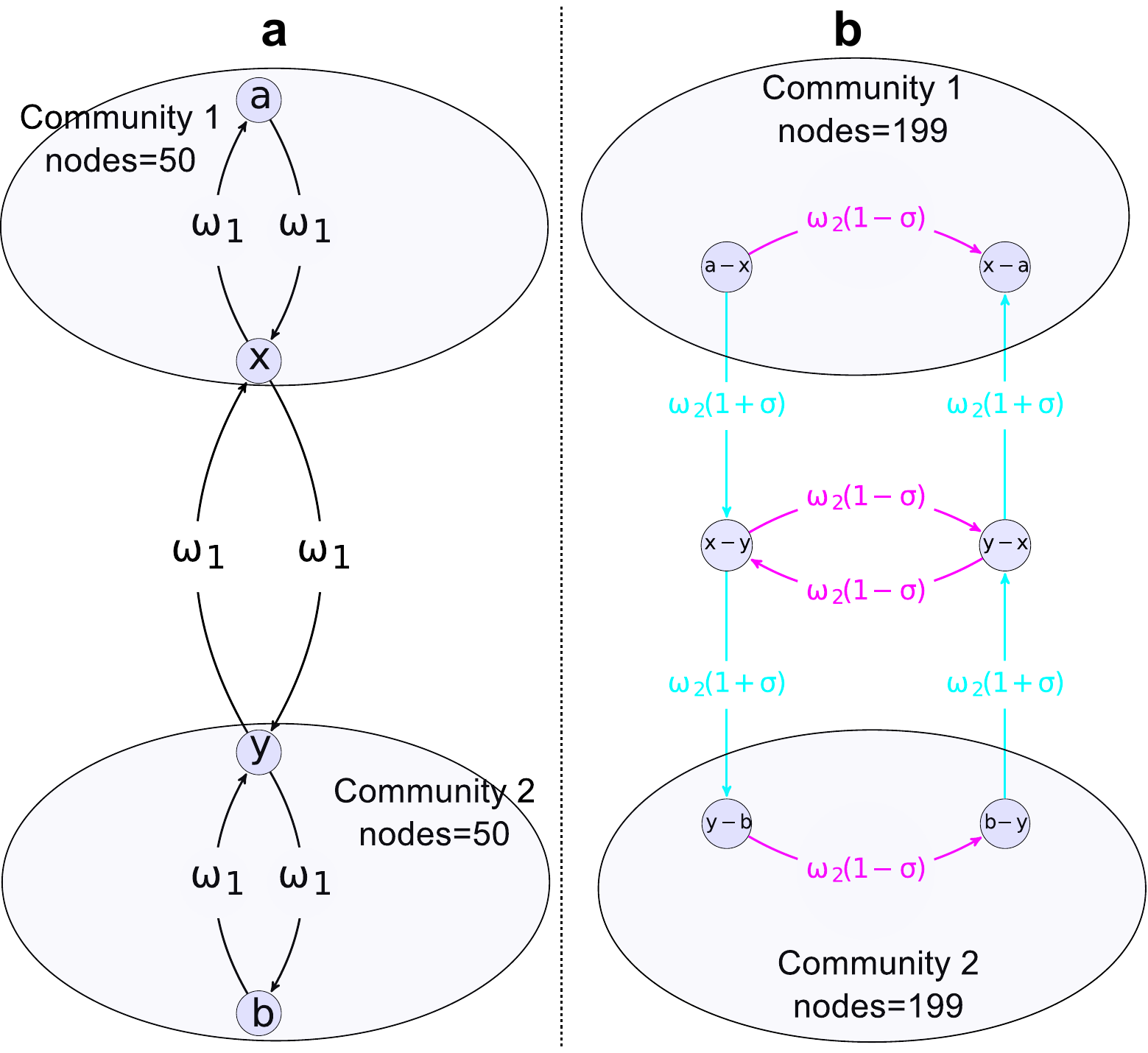}
 \caption{{\bf Schematic representation of our model for non-Markovian temporal networks} (a) The first-order aggregate network $G^{(1)}$ consists of two pronounced communities connected by directed inter-community links $(x,y)$ and $(y,x)$. (b) Weights in the corresponding second-order aggregate network $G^{(2)}$ are changed by means of a parameter $\sigma$. Positive values for $\sigma$ enforce two-paths across communities (turquoise) and inhibit two-paths within communities (magenta).}
 \label{fig:model}
\end{figure}

For this first-order network $G^{(1)}$, we construct a second-order network $G^{(2)}$ corresponding to Markovian edge activations as shown in panel (b) of Fig.~\ref{fig:model}.
Since $G^{(1)}$ has $400$ edges, $G^{(2)}$ has $400$ nodes, each corresponding to a directed edge in the first-order network.
As weights in the second-order network $G^{(2)}$, we consider a uniform constant $\omega_2$ which corresponds to a Markovian case in which consecutive edge activations are independently drawn.
We use the following simple strategy to introduce non-Markovian properties.
We first identify all edges $(x,y)$ that interconnect the two communities, i.e. where $x$ is a node in community 1 and $y$ is a node in community 2.
For these edges, we then identify two nodes $a,b$ such that $a$ is a node in community 1 adjacent to $x$ and $b$ is a node in community 2 adjacent to node $y$.
The basic idea of the model is to change the weights of those two-paths that involve edges $(a,x), (x,y), (y,x)$ and $(x,a)$.
The statistics of these two-paths is captured by the weights of \emph{edges} connecting \emph{nodes} $(a,x)$, $(x,y)$, $(x,a)$, $(a,x)$ in the second-order network (see panel (b) in Supplementary Fig.~\ref{fig:model}).

Based on a parameter $\sigma \in (-1,1)$, the weights of the second-order edges $(a,x) \rightarrow (x,y)$ and $(y,x) \rightarrow (x,a)$ are set to $\omega_2(1+\sigma)$, while the weights of second-order edges $(a,x) \rightarrow (x,a)$ and $(y,x) \rightarrow (x,y)$ are set to $\omega_2(1-\sigma)$.
Weights of second-order edges including the nodes $b$ and $y$ are adjusted analogously (see panel (b) in Supplementary Fig.~\ref{fig:model}).
By this means, positive values for $\sigma$ increase the weights of two-paths \emph{across communities} at the expense of two-paths \emph{within communities}.
Negative values for $\sigma$ increase the weights of two-paths \emph{within communities} at the expense of two-paths \emph{across communities}.
A value of $\sigma=0$ yields a second-order aggregate network with uniform weights $\omega_2$ which - by construction - corresponds to a Markovian case.

For $\sigma\neq0$, the above procedure leads to transition matrices $\mathbf{T}^{(2)}\neq \mathbf{\tilde{T}}^{(2)}$ which are however consistent with the same weighted aggregate network $G^{(1)}$.
This can be confirmed by checking that for all $\sigma \in (-1,1)$, the stationary activation frequencies of edges captured by the leading eigenvector $\boldsymbol\pi$ of $\mathbf{T}^{(2)}$ are the same.
The change of second-order weights by our model imply
\begin{align*}
 T^{(2)}_{(a,x)(x,y)}=\omega_2(1+\sigma),&\quad T^{(2)}_{(y,x)(x,a)}=\omega_2(1+\sigma)\, ,\\
 T^{(2)}_{(a,x)(x,a)}=\omega_2(1-\sigma),&\quad T^{(2)}_{(y,x)(x,y)}=\omega_2(1-\sigma)\,.
\end{align*}
Since the $j$-th component of the stationary distribution of the second-order network is given by $\left(\boldsymbol\pi\right)_j=\sum_{i}\left(\boldsymbol\pi\right)_i T^{(2)}_{ij}$ the changes above only influence entries $\left(\boldsymbol\pi\right)_{(x,a)}$ and $\left(\boldsymbol\pi\right)_{(x,y)}$ in the leading eigenvector of $\mathbf{T}^{(2)}$.
Let $\tilde{\boldsymbol\pi}=\tilde{\boldsymbol\pi}\mathbf{\tilde{T}}^{(2)}$ and $\boldsymbol\pi=\boldsymbol\pi\mathbf{T}^{(2)}$.
Then for an entry $\left(\boldsymbol\pi\right)_{(x,a)}$ we can write
\begin{align*}
\left(\boldsymbol\pi\right)_{(x,a)}=&\sum_{i}\left(\boldsymbol\pi\right)_{(i,x)} T^{(2)}_{(i,x)(x,a)}\\
=&\sum_{i\neq a,y}\left(\left(\boldsymbol\pi\right)_{(i,x)} T^{(2)}_{(i,x)(x,a)}\right)\\
&+\left(\boldsymbol\pi\right)_{(a,x)}T^{(2)}_{(a,x)(x,a)}+\left(\boldsymbol\pi\right)_{(y,x)}T^{(2)}_{(y,x)(x,a)}\,.
\end{align*}
Recall that our transformations only change the entries for $(x,a)$ and $(x,y)$ therefore it holds that $\left(\boldsymbol\pi\right)_{(i,x)}=\left(\tilde{\boldsymbol\pi}\right)_{(i,x)}$ for all $i$.
This yields
\begin{align*}
\left(\boldsymbol\pi\right)_{(x,a)}=&\sum_{i\neq a,y}\left(\left(\tilde{\boldsymbol\pi}\right)_{(i,x)} T^{(2)}_{(i,x)(x,a)}\right)\\
&+\left(\tilde{\boldsymbol\pi}\right)_{(a,x)}T^{(2)}_{(a,x)(x,a)}+\left(\tilde{\boldsymbol\pi}\right)_{(y,x)}T^{(2)}_{(y,x)(x,a)}\,.
\end{align*}
Furthermore, we can plug in the definitions for $\mathbf{T}^{(2)}$ from above and also use that $T^{(2)}_{(i,x)(x,a)}=\tilde{T}^{(2)}_{(i,x)(x,a)}$ for all $i \notin \{a, y\}$.
\begin{align*}
\left(\boldsymbol\pi\right)_{(x,a)} = & \sum_{i\neq a,y}\left(\left(\tilde{\boldsymbol\pi}\right)_{(i,x)} \tilde{T}^{(2)}_{(i,x)(x,a)}\right) + \left(\tilde{\boldsymbol\pi}\right)_{(a,x)}\omega_2(1-\sigma)\\
            & + \left(\tilde{\boldsymbol\pi}\right)_{(y,x)}\omega_2(1+\sigma)\\
            =& \sum_{i\neq a,y}\left(\left(\tilde{\boldsymbol\pi}\right)_{(i,x)} \tilde{T}^{(2)}_{(i,x)(x,a)}\right) \\
            & +\left(\tilde{\boldsymbol\pi}\right)_{(a,x)}\omega_2-\left(\tilde{\boldsymbol\pi}\right)_{(a,x)}\omega_2\sigma\\
            & + \left(\tilde{\boldsymbol\pi}\right)_{(y,x)}\omega_2+\left(\tilde{\boldsymbol\pi}\right)_{(y,x)}\omega_2\sigma\,.
\end{align*}
Since $\mathbf{\tilde{T}}^{(2)}$ is built from a regular graph it holds that $\omega_2=\tilde{T}^{(2)}_{(i,x)(x,a)}$ for all $i$. Hence,
\begin{align*}
\left(\boldsymbol\pi\right)_{(x,a)} = & \sum_{i\neq a,y}\left(\left(\tilde{\boldsymbol\pi}\right)_{(i,x)} \tilde{T}^{(2)}_{(i,x)(x,a)}\right) + \left(\tilde{\boldsymbol\pi}\right)_{(a,x)}\tilde{T}^{(2)}_{(a,x)(x,a)}\\
&-\left(\tilde{\boldsymbol\pi}\right)_{(a,x)}\omega_2\sigma+\left(\tilde{\boldsymbol\pi}\right)_{(y,x)}\tilde{T}^{(2)}_{(y,x)(x,a)}+\left(\tilde{\boldsymbol\pi}\right)_{(y,x)}\omega_2\sigma\\
=&\sum_i\left(\left(\tilde{\boldsymbol\pi}\right)_{(i,x)} \tilde{T}^{(2)}_{(i,x)(x,a)}\right)\\
&-\left(\tilde{\boldsymbol\pi}\right)_{(a,x)}\omega_2\sigma+\left(\tilde{\boldsymbol\pi}\right)_{(y,x)}\omega_2\sigma\\
=&\left(\tilde{\boldsymbol\pi}\right)_{(x,a)}-\left(\tilde{\boldsymbol\pi}\right)_{(a,x)}\omega_2\sigma+\left(\tilde{\boldsymbol\pi}\right)_{(y,x)}\omega_2\sigma\\
=&\left(\tilde{\boldsymbol\pi}\right)_{(x,a)}\,.
\end{align*}
In the last step we use that the stationary distribution $\tilde{\boldsymbol\pi}$ is uniform and thus $\left(\tilde{\boldsymbol\pi}\right)_{(a,x)} = \left(\tilde{\boldsymbol\pi}\right)_{(y,x)}$.
From an analogous argumentation, we can derive $\left(\boldsymbol\pi\right)_{(x,y)}=\left(\tilde{\boldsymbol\pi}\right)_{(x,y)}$.
We thus confirm that $\boldsymbol\pi=\tilde{\boldsymbol\pi}$ and the stationary distribution is preserved for $\sigma \in (-1,1)$.
We finally refer the reader to a related model for non-Markovian temporal networks, which has been introduced very recently, during the revision of our manuscript~\cite{Lambiotte2014}.
Different from our approach, the model introduced in this recent work generates realisations that do not preserve a given weighted aggregate network, which however is the particular focus of our approach.

\newpage

\renewcommand{\refname}{Supplementary References}

\end{document}